\documentclass{PoS}
\usepackage{graphicx}

\title{Free energy of static quarks and the renormalized Polyakov loop in full QCD}

\ShortTitle{Free energy of static quarks}

\author{\speaker{Konstantin Petrov}, RBC-Bielefeld collaboration\\
        NIC, DESY Zeuthen, Platanenallee 6, 15738 Zeuthen, Germany\\
        E-mail: \email{Konstantin.Petrov@desy.de}}

\abstract{
We present results from a detailed study of singlet free energies  
in full QCD with realistic quark masses. 
An improved scheme for the non-perturbative renormalization of the Polyakov 
loop is used and we compare its temperature dependence for QCD with
different flavor content. We also analyze screening masses extracted from 
singlet free energies at various temperatures close to  and above the
QCD transition temperature. We conclude that the temperature dependence of
screening masses is well described by perturbation theory up to a 
non-perturbative pre-factor. An effective running coupling has been determined
for all temperature values giving additional insight into screening phenomena
at high temperature.
}

\FullConference{The XXV International Symposium on Lattice Field Theory\\
		 July 30-4 August 2007\\
		 Regensburg, Germany}

\begin{document}

\section{Introduction} 
One of the most important properties of the quark gluon plasma is the screening of color 
charges \cite{mclerran}. 
It is expected that quarkonium states are dissociated due to screening of the color charges at temperatures 
somewhat higher than the deconfinement temperature. 
Moreover, Matsui and Satz conjectured \cite{satz} that quarkonium suppression 
can be viewed as a signature for the formation of a Quark-Gluon Plasma (QGP).  
While a calculation of spectral functions is best suited to study this 
phenomenon, their reliable calculation at finite temperature
turns out to be quite difficult (see e.g. \cite{jacovac}). 
We present here a non-perturbative study of the screening of 
heavy quark-anti-quark interactions at finite temperature which is 
based on an analysis of the free energy of a static quark anti-quark 
pair \cite{mclerran}. Results for the free energy and its derivatives, 
the internal energy and entropy, can be used as 
input to potential model calculations, see e.g.\cite{Digal:2001iu,agnes07}.
Furthermore, the study of static free energies is interesting as it allows 
for a non-perturbative renormalization of Polyakov loops. 
While the latter is not an order parameter in the
presence of dynamical quarks it shows rapid variation in the transition 
region and therefore is widely used to describe
the transition (crossover) in full QCD, e.g. through effective mean-field 
theories. 

\section{Numerical Analysis}

This work is based on our large-scale finite temperature lattice calculations 
in (2+1)-flavor QCD in the region of  small quark masses \cite{Cheng:2007jq}. 
Our simulation have been performed with a  physical strange quark mass and 
degenerate light quark masses, $m_q=0.1m_s$ ($m_s$ 
being the strange quark mass), which correspond to  a light pseudo-scalar
mass $m_\pi\simeq 220MeV$. 
Lattice sizes vary between  $16^3 \times 4$ and $24^3 \times 6$ lattices 
which correspond to the same physical volume.   
The lattice spacing and thus the temperature scale has been fixed using the 
Sommer scale $r_0=0.469$ fm \cite{gray}. For every value of the
finite temperature coupling constant we performed corresponding zero
temperature simulations where we extracted the zero temperature static
quark potential. This has been used to set the temperature scale and to
determine the renormalization constants used for a renormalization of the
finite temperature free energies. 

In our simulations we used the exact RHMC algorithm for simulations of 
(2+1)-flavor QCD. Further details about our simulations can be found in 
Refs.\cite{Cheng:2007jq,us}. 
On a large set of gauge field configurations, which are separated by 10 trajectories, 
we have calculated the singlet free energy, $F_1(r,T)$, of a static
quark anti-quark pair. The singlet free energy, as well as the 
zero temperature static potential, are defined up to an additive 
constant $C$. Using temporal Wilson lines  $W(\vec{r})$ we write the former 
as 
\begin{equation}
\exp(-F_1(r,T)/T+C)= \frac{1}{3} \langle Tr W(\vec{r}) W^{\dagger}(0) \rangle 
\;\;, \;\;   C=2 N_{\tau} \ln(Z_R) \;.
\label{f1def}
\end{equation}
The zero temperature potential is calculated from smeared Wilson 
loops  $W(r,\tau)$, 
\begin{equation}
\nonumber
V_{(T=0)}(r)=-
\lim_{\tau\rightarrow\infty}
\ln \left( Z_{R}(\beta)^2\frac{W(r,\tau)}{W(r,\tau+1)} \right) \; .
\label{eq:Vren}
\end{equation}
Here $Z_R$ is zero temperature renormalization constant. 
%$a$ - temperature dependent lattice spacing. 
Some details of our Wilson loop calculations are 
discussed in \cite{us}.
The above definition of a singlet free energy requires gauge fixing. For 
all our calculations we use Coulomb gauge 
as it is done also in many previous closely related studies 
\cite{ophil02,okacz02,digal03,okacz04,petrov04,okacz05}.
In the zero temperature limit the singlet free energy coincides with the 
well-known static potential and, in fact, an analysis in a fixed gauge
has been used also in this limit to 
calculate it  \cite{milc04}. At finite temperature $F_1(r,T)$ characterizes
the in-medium modification of inter-quark forces and color screening. 

The renormalization constant $Z_R$ introduced in Eq.\ref{eq:Vren} is 
determined by matching result for the zero temperature static potential
to the string potential, $V_{string}(r)=- \pi/12r+\sigma r$, at 
distance $1.5r_0$. Renormalizing the potentials
at one point makes results obtained for a wide range of cut-off values
coincide quite well at all distances \cite{Cheng:2007jq}.
Unlike in our earlier work, where we renormalized potentials at a short
distance point, we now do so at a larger distance, $r=1.5r_0$
This has the advantage that we can use the same large distance string
potential, $V_{string}(r)$, for all our potentials, irrespective
of the flavor content used in the simulation.
At short distances this would not be the case as the running of the 
coupling entering the Coulomb term of the potential is sensitive to the
number of flavors \cite{Cheng:2007jq}. 
%In our earlier analysis we used a simple Cornell-like parametrisation of the
%zero temperature potential at all values of coupling. This ignored 
%effects arising from a running of the coupling. 
%As we now have detailed information on the potential at zero temperature 
%for every value of the finite-temperature coupling, {\it i.e.} at each value
%of the cut-off,  we can avoid this problem completely. 
Of course, it also is important to note that we now have sufficient 
statistics to perform a reliable matching of potentials at the
relatively large distance  $r=1.5r_0$. This also has the advantage that
lattice-induced short-distance artifacts are significantly reduced in our
analysis. 
%However, while being more rigerous this procedure should differ 
%little from the old one as due to the use of improved action
%lattice artefacts seem to be significantly supressed.

\begin{figure}
\includegraphics[width=0.7\textwidth]{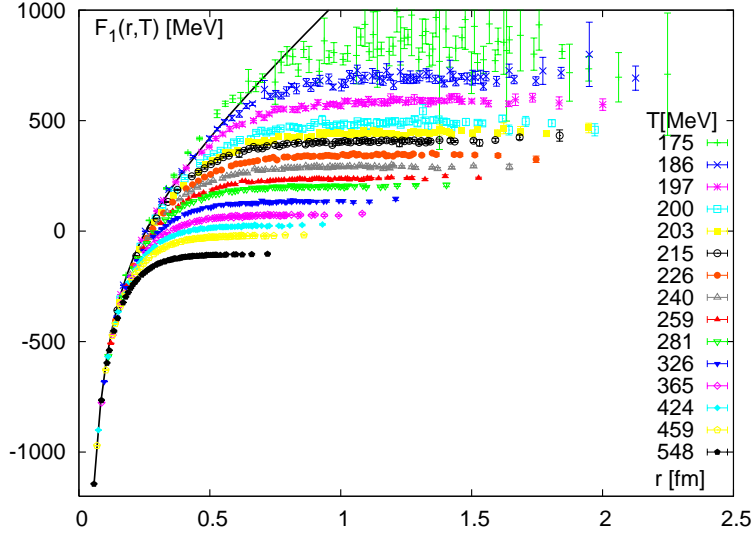}
\caption{The singlet free energy calculated in (2+1)-flavor QCD 
on $24^3 \times 6$ lattices at various 
temperatures. The solid line shows a fit to the zero temperature potential.}
\label{fig:f1}

\end{figure}

In Fig. \ref{fig:f1} we show the singlet free energy calculated in
$(2+1)$-flavor QCD on  
$24^3 \times 6$ lattices together with the $T=0$ static potential. As one 
can see from the figure, $F_1(r,T)$ is temperature independent at small 
distances and coincides with the zero temperature potential
as expected. At large distances the singlet free energy approaches a 
constant value. This can be related to string breaking at low temperature 
and color screening at high temperatures. Note that the distance at which 
the free energy effectively flattens off is decreasing with increasing 
temperature. This is another indication for color screening at high temperature.

\section{Renormalized Polyakov Loop}

The expectation value of the Polyakov loop $\langle L(\vec{r}) \rangle =\langle Tr W(\vec{r}) \rangle$ 
is the order parameter
for deconfinement in pure gauge theories. In full
QCD there is no local order parameter because dynamical quarks break the relevant $Z(3)$ symmetry explicitly.
Still it remains an interesting quantity that can  be used 
to analyze deconfinement in QCD as 
it shows a rapid increase in the
crossover region \cite{us,milcthermo,fodor06} and can be used to determine 
the transition temperature \cite{us,fodor06}.
The Polyakov loop defined above strongly depends on the lattice spacing and requires  renormalization.
The singlet free energy and the Polyakov loop correlator, which defines the color averaged free energy, satisfy the cluster decomposition
\begin{equation}
  \lim_{r \rightarrow \infty}   \exp(-F_1(r,T)+C)=\frac{1}{9} \lim_{r \rightarrow \infty}
\langle L(\vec{r}) L^{\dagger}(0) \rangle = |\langle L (0)\rangle|^2
\equiv L^2.
\end{equation}
The normalization constant in the above expression is fixed through the 
renormalization of the zero temperature potential
as discussed in the previous section. Thus, at large distances  
color averaged and singlet free energies approach the same 
constant, $F_{\infty}(T)$.
We defined the renormalized Polyakov loop as 
\begin{equation}
L^{ren}(T)=\exp(-\frac{F_{\infty}(T)}{2 T})  \; ,
\label{eq:lren1}
\end{equation}
which due to the cluster decomposition is the same as 
\begin{equation}
L^{ren}(T)=(Z_R(g^2))^{N_{\tau}}L \; .
\label{eq:lren2}
\end{equation}
Here the renormalization constants are the same as in Eq.\ref{eq:Vren}.
\begin{figure}
\includegraphics[width=0.6\textwidth]{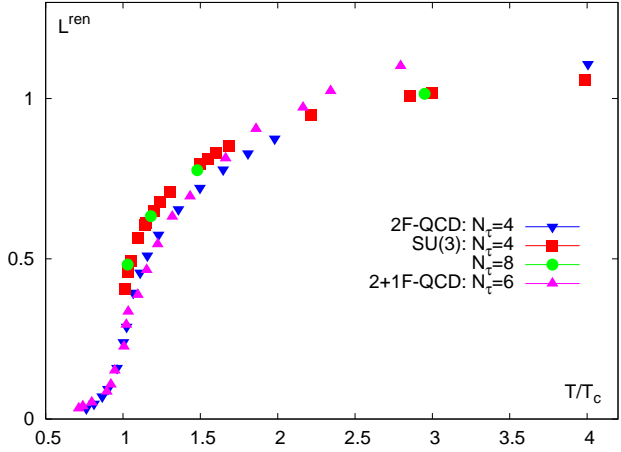}
\caption{
The renormalized Polyakov Loop calculated in ($2+1$)-flavor QCD
on $16^3\times 4$ and $24^3 \times 6$ lattices
for different quark masses along a line of constant physics. 
Also shown in the figure are corresponding 
result from calculations in 2-flavor QCD and the SU(3) gauge theory.}
\label{fig:lren_sum}
\end{figure}
Our numerical results for $L^{ren}(T)$ obtained in simulations with different 
number of light quark flavors (two-flavour and pure gauge see \cite{okacz05}) 
and different lattice spacings are summarized in Fig.\ref{fig:lren_sum}. One 
can see from that figure that $L^{ren}(T)$ shows an almost 
universal behavior as function of $T/T_c$ for all quark masses studied by us, 
including the 2-flavor simulation. 
This suggests, that in the region of small quark masses, which has been
studied by us, the flavor and quark mass dependence of the deconfinement 
transition can be almost entirely understood in terms
of the flavor and quark mass dependence of the transition temperature $T_c$. 
We note that a similar renormalization procedure of the Polyakov loop as
discussed above 
%and which is based only on the zero
%temperature potential was first 
was also used in Ref. \cite{fodor06}. 
%We find, however, that the results do not change significantly
%compared to the previous procedure, where the singlet free energy was 
%normalized to a Cornell potential at short distances \cite{petrov06}.
We also note that in Ref. \cite{fodor06} 
a different normalization convention
for the zero temperature potential has been used and therefore the absolute 
value of the Polyakov loop is quite different. 

\section{Running Coupling}
Now let us discuss our results on the distance and temperature dependence of 
the singlet free energies in terms of a running coupling constant. We define 
the temperature-dependent coupling analogously to the zero temperature case 
through the free energy,

\begin{equation}
\alpha_{eff}(r,T)=\frac{3}{4}r^2\frac{dF_1(r,T)}{dr}  \; .
\end{equation}

\begin{figure}
\hspace*{-0.8cm}
\includegraphics[width=0.53\textwidth]{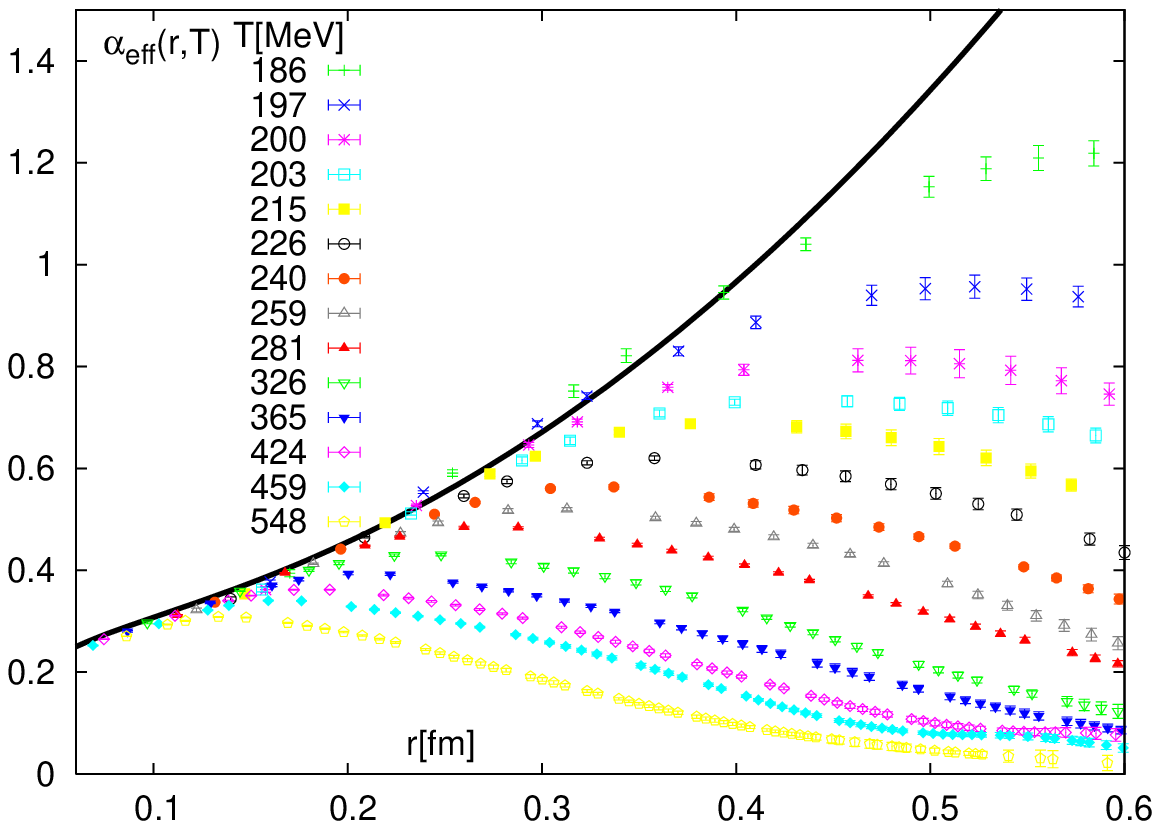}
\includegraphics[width=0.53\textwidth]{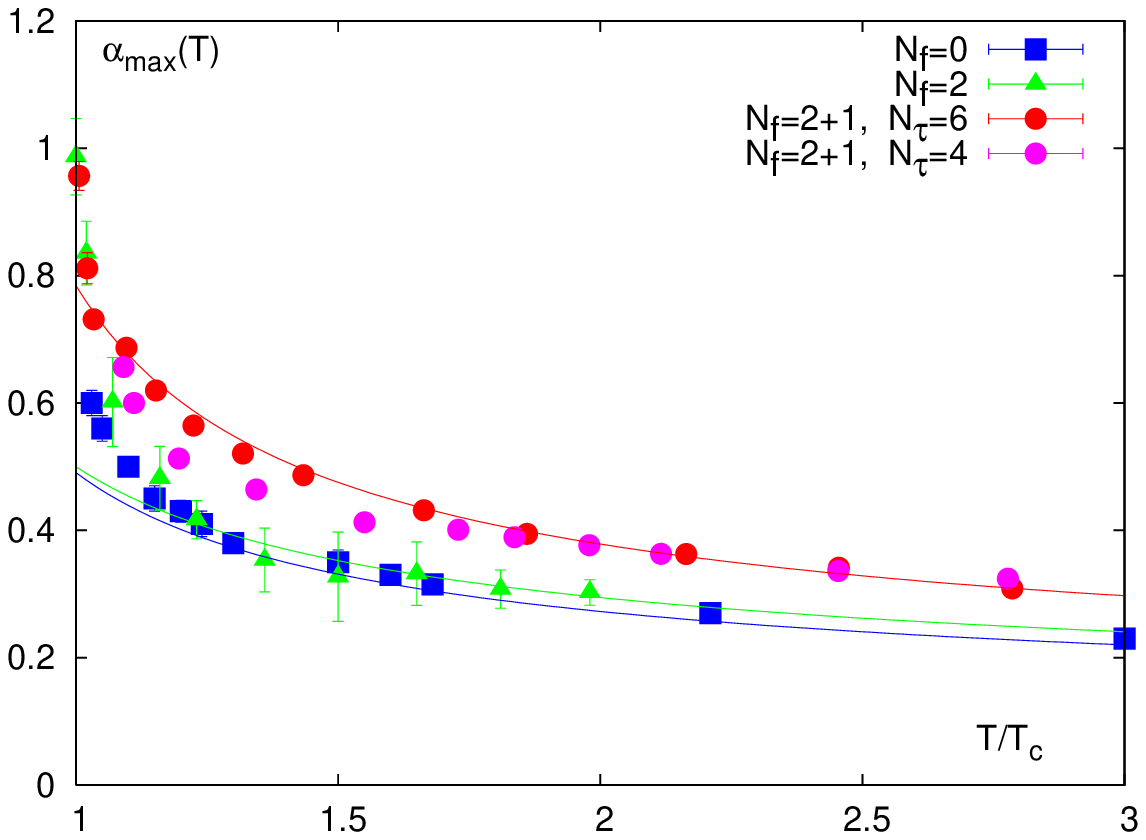}
\caption{Temperature dependence of the effective running coupling at 
various values of the temperature (left). The black solid line shows the
result at zero temperature. The right hand figure shows the maximal value
attained by $\alpha_{eff}(r,T)$ in calculations of QCD with different flavor
content.} 
\label{alpha}
\end{figure}

In Fig.\ref{alpha} (left) we present our results for the running coupling.  
At short distances, especially at low temperatures, it coincides with the zero 
temperature coupling until screening sets in. After reaching a maximal value 
it decreases slowly. At higher temperature the shift of the maximum to 
shorter and shorter distances again indicates that screening effects set in
at shorter distances. However still the running coupling rises quadratically
at intermediate distances, before screening sets in. This is arises from
the linear confinement part of the static potential and signals the remnants
of confining forces even at rather high temperatures. We, therefore, can 
conclude that remnants of the confining force still are important for physics
at moderately large distances even in the high temperature phase of QCD. 
The maximal value of the effective running coupling along with fits inspired 
by a  perturbative Ansatz are shown in the right hand part of Fig.~\ref{alpha}.
One can see that the maximal
value decreases only slowly with temperature and even at three times the
transition temperatures it is only a factor five smaller than at 
the transition temperature. 
For comparison we show here also results obtained in simulations
of 2-flavor QCD with larger quark mass ($m/T=0.4$) and pure gauge theory 
results. The large value of the effective coupling is again due to the 
linearly rising term in the potential. As one can see from the comparison of 
results obtained for two different lattice spacings, cut-off effects for the 
effective coupling are small.

\section{Screening masses}
To analyze the exponential screening of free energies at large distances
we extract screening masses. They are determined from fits to the large
distance part of the free energy where we subtracted the asymptotic cluster 
value,
\begin{equation}
F_1(r,T)-F_1(r=\infty,T)=-\frac{4}{3}\frac{\alpha(T)}{r}\exp(-m_D(T)r) \ .
\end{equation}
To leading order in perturbation theory the screening (Debye) mass is
given by
\begin{equation}
\frac{m_D}{T}=A\left( 1+\frac{N_f}{6} \right)^{1/2}g(T)  \; .
\end{equation}
\begin{figure}
\includegraphics[width=0.6\textwidth]{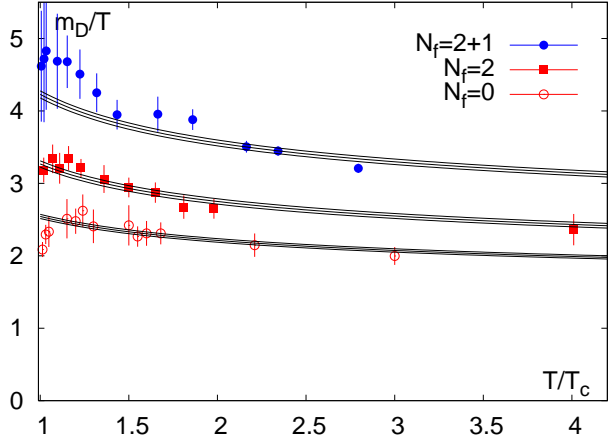}
\caption{Screening masses in ($2+1)$)-flavor QCD versus temperature. 
For comparison 
results from calculations in 2-flavor QCD and pure gauge theory 
are included. Lines represent fits to leading order perturbation theory 
with free pre-factor $A$.}
\label{fig:screen1}
\end{figure}
In Fig.\ref{fig:screen1} we show screening masses as function of temperature. 
In addition to results from our $(2+1)$-flavor simulations we also show 
results from calculations 
performed in 2-flavor QCD and pure gauge theory. These numerical results are compared 
to leading order perturbation theory, where we allow for a pre-factor $A$ 
which is fixed by a fit to the data. This pre-factor is
slightly different for each case. We can see, however, that the 
temperature dependence of the screening masses in all cases 
is described quite well by the perturbative Ansatz.

\section{Conclusions}
We have significantly expanded our analysis of heavy-quark related physics 
through large scale simulations in (2+1)-flavor QCD. 
%While confirming 
%previous results and predictions we introduced improved method of 
%renormalizing Polyakov loop which takes into account the running of the 
%coupling and is less dependent on short-distance physics. 
The renormalized Polyakov loop shows little cut-off dependence and can be 
calculated reliably on relatively coarse lattices. We find that its
flavor and quark mass dependence can be absorbed
almost entirely in the flavor and quark mass dependence of the transition 
temperature. The analysis of screening masses shows that non-perturbative 
effects can be well absorbed in a pre-factor in front of the leading order 
perturbative result which is temperature independent in the entire
temperature range analyzed here. This indicates that the true perturbative 
limit, corresponding to $A\equiv 1$, 
is approached only very slowly.

The temperature dependent effective running coupling constant rises 
quadratically  at moderate distances also above the transition temperature.
After reaching a maximal value that is reached at smaller distances with increasing
temperature it drops exponentially. Also this maximal value decreases 
only slowly with temperature.

\end{document}